*Review*

# Unlocking the Full Potential of Taylor Dispersion Analysis Through Advanced Detection

**Rami Laoufi and Agnès Hagège ***

Université de Lyon, CNRS, Université Claude Bernard Lyon 1, Institut des Sciences Analytiques, UMR 5280, 69100 Villeurbanne, France; agnes.hagege@isa-lyon.fr
* Correspondence: agnes.hagege@isa-lyon.fr

**Abstract**

Today, pharmaceutical sciences rely on a wide range of innovative therapeutics of varying sizes: nanoparticles, proteins, antibodies, and nanocarriers for drug delivery. Measuring their size is one of the crucial parameters in their characterization, to ensure drug safety, stability, and effectiveness. Over the past two decades, Taylor Dispersion Analysis has emerged as an important sizing method. Its scope of application is perfectly suited to the sizes of these therapeutics. By design, it can be coupled with a wide range of detectors, and extensive research has been conducted in this field. In this review, we will present an overview of the studies conducted, and discuss how the application drives the choice of one detector or another (size measurement, aggregation studies, interaction studies, and other analyses, performed in both simple biological buffers and in more complex media, particularly serum) and makes Taylor Dispersion Analysis a highly versatile tool.

## 1. Introduction

The development of new treatments has led to an increasingly diverse range of solutions, both in terms of therapeutics themselves and drug delivery systems. In addition to traditional chemical compounds, modern drug development encompasses peptides and proteins, monoclonal antibodies, antibody–drug conjugates, nucleic acid-based therapeutics such as oligonucleotides and mRNA, and metallic nanoparticles. This diversity is further expanded by prodrugs [1], which are designed to undergo biotransformation to release their active form, and by drug delivery systems [2,3], including lipid nanoparticles, polymeric nanoparticles, and liposomes, developed to improve the stability, targeting, or bioavailability of therapeutic agents.

It is now generally admitted that size is a parameter of fundamental importance in biopharmaceuticals sciences. The size of an object refers to various concepts. In particular, we distinguish between (i) the "actual" size of the object—that is, the space it physically occupies, generally measured using transmission electron microscopy (TEM), and (ii) the effective size of the object in a given medium, known as the hydrodynamic size. The hydrodynamic size is often considered the most relevant size in biological media. Indeed, this size can be significantly different, depending in particular on the solvation in a given environment.

For any therapeutic treatment (drug or carrier), size determines its ability to cross the biological barriers and therefore its biodistribution and renal clearance ability. This factor has become increasingly important with the introduction of nanoparticles in the medical field. Although small size is important for safety reasons, it turns out that fast clearance can be detrimental, since it limits the time available to bind to and neutralize the targets. Size measurement can be used to monitor the stability of biopharmaceuticals (e.g., aggregation, protein unfolding). As an example, protein aggregation has to be monitored at all stages of production, since aggregates may elicit immune reactions and affect the binding specificity of antibody–drug conjugates. It is also particularly useful for verifying the drug load in drug delivery systems or for monitoring their degradation [4].

In addition to the measurement of the size of compounds per se, numerous studies have used variations in this size to determine affinity constants (e.g., drug/antibody).

In recent years, Taylor Dispersion Analysis (TDA) has emerged as a method of choice among sizing methods, and the number of papers demonstrating its value in this field continues to grow. Its advantages and synergies with other techniques have been documented by Moser et al. [5]. Among them are very low sample consumption (in the nL range), the absence of bias toward larger analytes in a mixture, and an absolute, calibration-free analysis. In addition, its ability to be coupled with a large number of detectors relatively easily makes it a very versatile tool. However, just as in capillary electrophoresis, detection is one of the critical aspects of TDA, as the limitations in sensitivity, inherent to each detection method, are further exacerbated by the very small volumes injected.

In this review, we aim to provide an overview of the detection methods used in TDA and their implementation. They can either be detections that can be performed directly on the capillary, known as "on-line detection," or detections at the capillary outlet, known as "in-line detection". Significant efforts have been made to combine TDA with various detection methods, such as UV, fluorescence, and mass spectrometry, to name just the most common ones. However, a direct comparison of their respective sensitivities remains challenging. Indeed, sensitivity depends not only on the nature of the compounds and the analytical conditions but also on the compounds' size, which may vary greatly depending on the studies. We also place particular emphasis on the aforementioned applications. This review does not claim to be exhaustive, but rather aims to show how the inherent characteristics of each detector influence the field of application in which TDA is used.

## 2. Materials and Methods

This review is based solely on articles published in scientific journals in a language accessible to the entire scientific community (i.e., English). Data collection was conducted between 1 June and 14 September 2026. The article selection strategy is illustrated in Figure 1. The initial phase of collection was conducted using the Web of Science database based on the free-text keywords "Taylor Dispersion Analysis" and "Flow-Induced Dispersion Analysis." The screening phase was then conducted using controlled descriptors to eliminate areas of application not relevant to Taylor Dispersion Analysis. Following this screening phase, the collected data were compiled with data available on commercial equipment manufacturers' websites (Malvern, Fidabio, and Nanoscalemetrix), and duplicates were removed. Application notes were not included in this study. The identification phase was conducted manually and independently by the two authors to eliminate articles whose concepts differ from those discussed in the review (e.g., "influence of Taylor diffusion on chromatographic profiles," "Taylor cone in electrospray"). The eligibility of articles was determined based on disjunctive eligibility criteria defined by the two authors, namely:

Criterion 1: The publication must address TDA detection at any level: unconventional detection methods, improvement in detection limits, etc.

Criterion 2: The publication must have a stated application in the biopharmaceutical field.

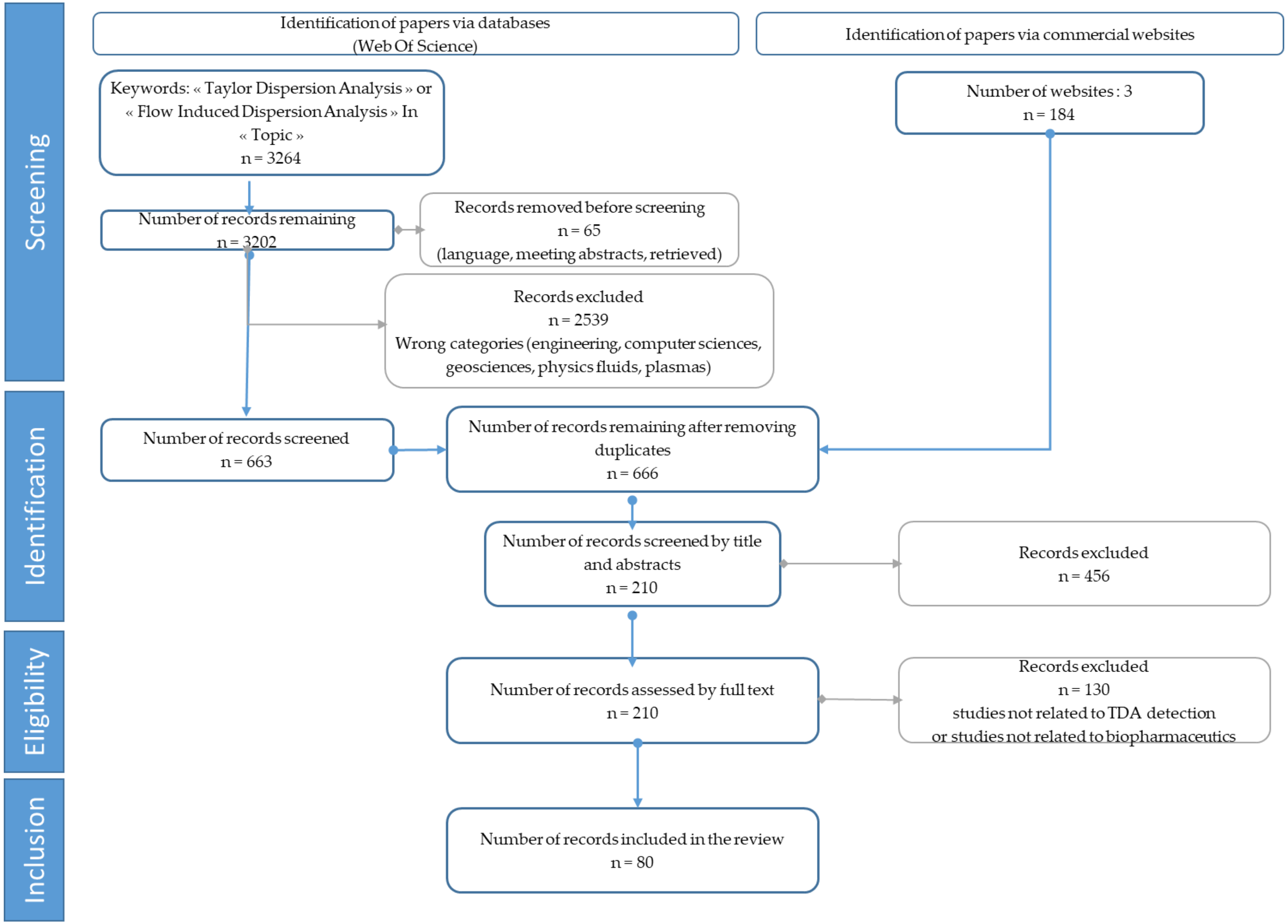

**Figure 1.** PRISMA flowchart illustrating the methodology used in the review.

At the conclusion of this screening phase, the articles selected by only one of the reviewers were discussed to determine whether they should be included.

## 3. Principle of TDA

Taylor Dispersion is a phenomenon first described by Sir Geoffrey Taylor in 1953 [6,7], concerning the dispersion experienced by a molecule subjected to laminar flow in a tube, when diffusion along the tube is negligible. The principle is illustrated in Figure 2.

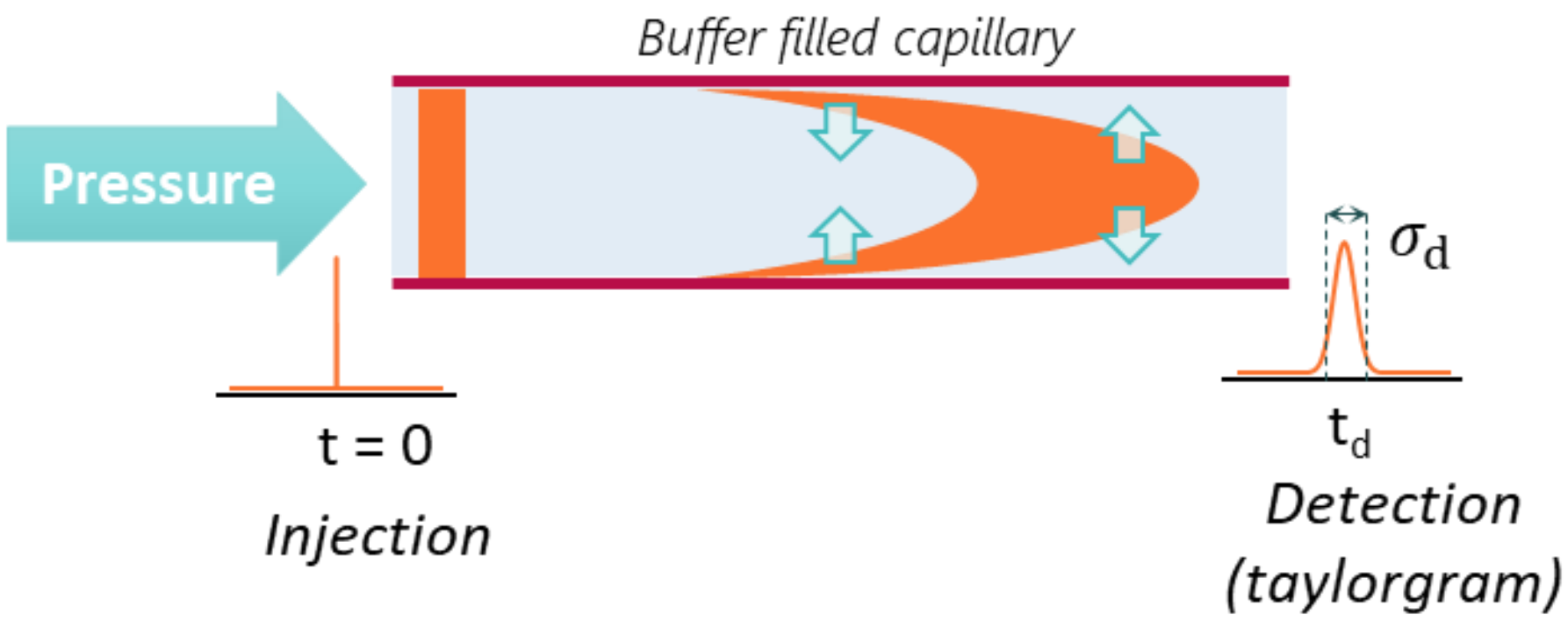

**Figure 2.** Principle of the method: Axial dispersion ($\sigma_d$) of molecules/nanoparticles can be related to their diffusion coefficient and, consequently, to their hydrodynamic radius.

A laminar flow in a tube with a circular cross-section has a parabolic velocity profile. The fluid velocity is highest at the center of the tube and nearly zero at the walls. When a sample is injected, the species at the center of the tube are then moving faster than species near the walls, and a concentration gradient develops (highlighted in orange). The analysis using Taylor Dispersion involves neglecting axial diffusion. The Peclet number Pe must be chosen according to Equation (1):

$$\mathrm{Pe} = \frac{\mathrm{uR_c}}{\mathrm{D}} \geq 40, \tag{1}$$

where u is the linear velocity of the fluid, $R_c$ the capillary diameter and D the molecular diffusion coefficient of the solute

The dimensionless characteristic diffusion time τ must be chosen according to Equation (2):

$$\tau = \frac{\mathrm{Dt_0}}{\mathrm{R_c^2}} \geq 2.5, \tag{2}$$

where $t_0$ is the residence time of the peak in the capillary.

Under these conditions, only radial diffusion has a significant effect. Species located at the center of the capillary diffuse toward the walls, causing their velocity to decrease. Similarly, species located at the walls diffuse toward the center and accelerate. These diffusion phenomena cause an oscillation in the velocity of the species in solution around the average velocity. For more information about the influence of experimental parameters, the reader could refer to [8].

By injecting a small plug of sample in a capillary and applying the adequate velocity, the signal detected at the capillary outlet can then be modeled as a Gaussian peak and is called a taylorgram. Small molecules diffuse rapidly, resulting in a small deviation around the average velocity, and thus around $t_0$. The detected peak is therefore narrow. In contrast, for a larger analyte, such as a protein or a nanoparticle, diffusion is slower, and the observed peak will be broad.

The peak of a taylorgram for a solution containing a single molecule with a diffusion coefficient D can be modeled by a Gaussian function, according to Equation (3):

$$S(t) = y_0 + \frac{A}{\sigma\sqrt{2\pi}} \exp - \frac{1}{2}\frac{(t-t_0)^2}{\sigma^2}, \quad (3)$$

where t is the experimental time, $t_0$ and σ are the mean time and temporal variance of the fitted Gaussian distribution, respectively, $y_0$ the baseline intensity, and A is a constant depending on the detection sensitivity and the concentration of the sample. For a sample containing several compounds of distinct sizes, the signal will correspond to the sum of several Gaussian distributions, whose characteristics $\sigma_i$ and $A_i$ correspond, respectively, to the response and the temporal variance of each compound i.

The diffusion coefficient of the solute can then be deduced from the parameters of the Gaussian function, according to Equation (4):

$$D = \frac{R_c^2 t_0}{24\sigma^2}, \quad (4)$$

where $R_c$ is the capillary radius.

The hydrodynamic radius is defined as the radius that a spherical particle would have, given its diffusion coefficient. The size of the analyzed compound can then easily be calculated using the Stokes–Einstein law, according to Equation (5).

$$D = \frac{k_B T}{6\pi\eta R_h}, \quad (5)$$

with D the diffusion coefficient, $k_B$ the Boltzmann constant, T the temperature, η the viscosity of the medium and $R_h$ the hydrodynamic radius of the solute.

Cipelletti et al. demonstrated the ability to access the full distribution of polydisperse samples via an adequate data processing of the taylorgrams (called Constrained Regularized Linear Inversion), based on an analogy to the analysis of cumulants performed in DLS [9].

Moreover, examining the variations in diffusion coefficients rather than the diffusion coefficient itself is of primary interest for the study of kinetically fast biomolecular interactions. This variant of TDA, named flow-induced dispersion analysis (FIDA), consists of injecting the small binding partner in media containing the larger biomolecule The diffusion coefficient shift becomes the hallmark of their interaction, as illustrated in Figure 3.

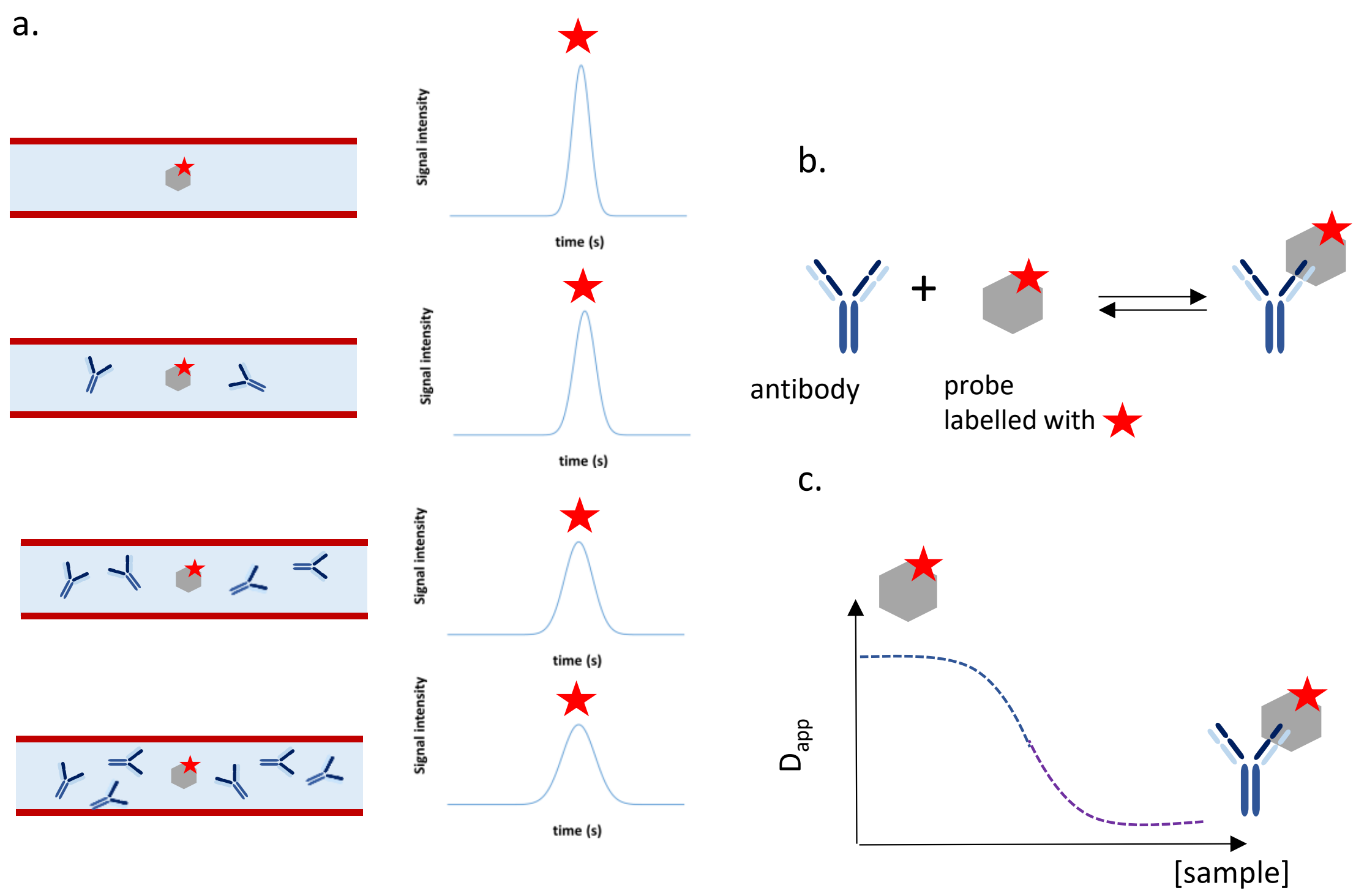

**Figure 3.** Principle of interaction measurement by FIDA, illustrated using the example of an antibody that recognizes a small probe. (**a**) The probe is injected into a sample with increasing concentrations; (**b**) the probe is recognized by the antibody; (**c**) the concentration of antibody can be deduced from the plot $D_{app}$ vs. sample concentration.

In this case, the measurement of the diffusion coefficient is actually an apparent diffusion coefficient ($D_{app}$), expressed according to Equation (6):

$$D_{app} = (1 - x)\ D_{unbound\ compound} + x\ D_{complex}, \quad (6)$$

where $D_{unbound\ compound}$, and $D_{complex}$ are the diffusion coefficients of the fluorescent probe and of the complex, respectively, and x is the fraction of fluorescent probe, which is complexed.

## 4. The Role of TDA in the Field of Size Measurement Methods

Size determination can be performed through an array of methods, some of which are complementary.

Various methods are size-based separation methods such as size-exclusion chromatography [10,11], or field-flow fractionation, especially asymmetric flow field-flow fractionation (AF4) [12,13]. Unlike size-exclusion chromatography, which is limited to objects no larger than 100–200 nm, field-flow fractionation has proven to be particularly useful for large objects (on the order of tens of µm). These techniques can be hyphenated with a wide range of detectors, UV and inductively coupled plasma–mass spectrometry (ICP-MS) being, in practice, among the most popular. However, the need for size standards is still a hurdle that limits the measurement of sizes. Moreover, the interactions with the stationary phase in SEC and sample loss due to interactions with the membrane are one of the main limitations.

Methods that determine size through the measurement of the diffusion coefficient are the most popular. Among these, we can mention dynamic light scattering (DLS), nuclear magnetic resonance (NMR), nanoparticle tracking analysis (NTA) fluorescence correlation spectroscopy (FCS) and fluorescence cross-correlation spectroscopy (FCCS), and more recently TDA. Figure 4 presents the applicable size range of these techniques.

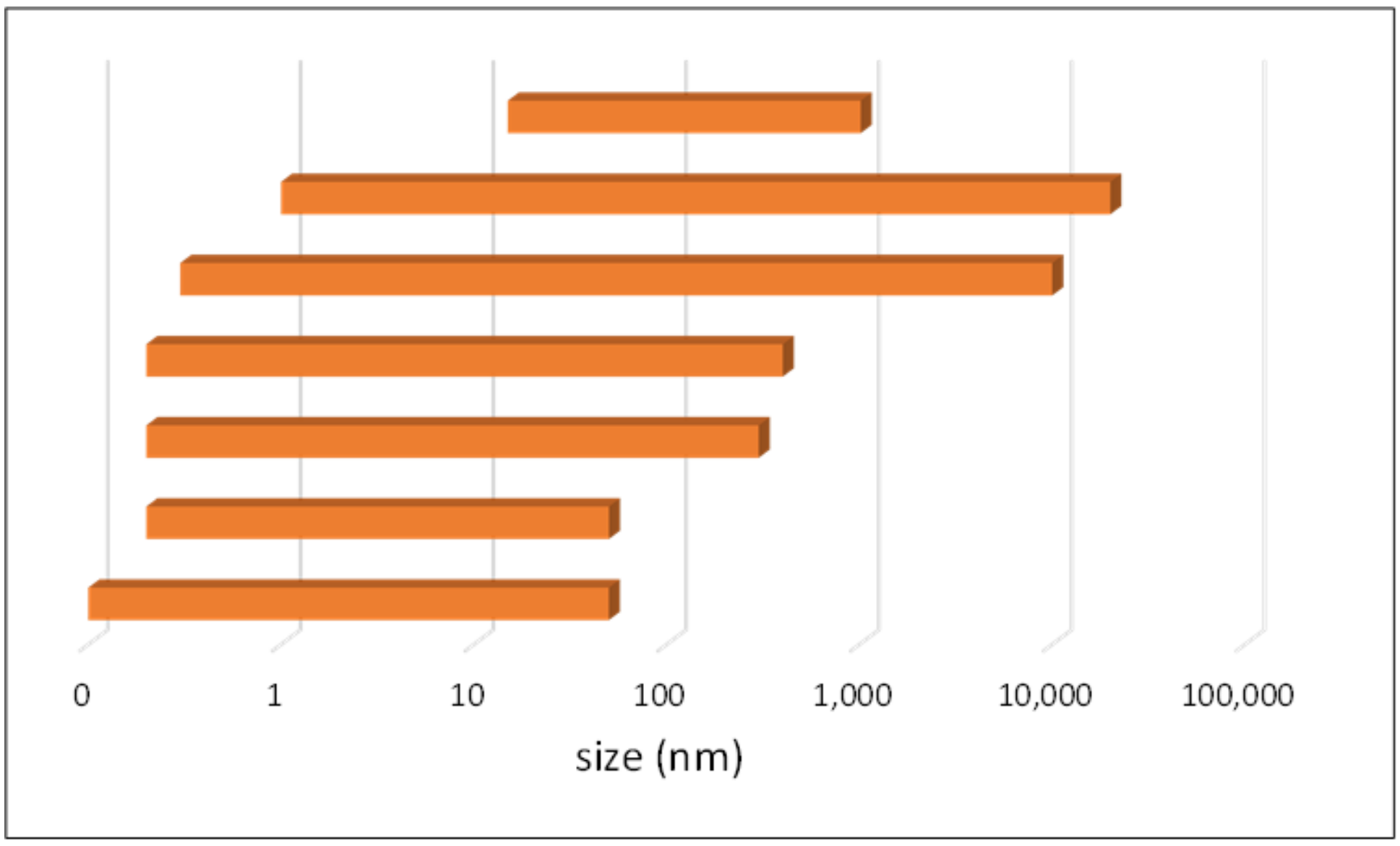

**Figure 4.** Comparison of the size range of the different techniques used for measuring hydrodynamic diameters.

DLS sizing is a robust and routine technique in aqueous media. It has been thoroughly validated [14], and commercial instruments and data processing software are widely available. However, the lack of specificity of the detection makes its use extremely challenging in biological media. The scattered intensity scales with $R_h^6$, which enhances the interferences caused by macromolecules, and the presence of large aggregates tends to bias the measured average hydrodynamic radius toward larger size and may hide the intrinsic solute signal.

NTA is also based on light scattering detection, but can track individual compounds, due to the use of a coupled device camera. It leads to an improvement in resolution compared to DLS. However, the range of sizes is more limited, especially when it comes to smaller sizes [15].

FCS and FCCS [16,17] measure the diffusion coefficient from the fluctuations of the fluorescence intensity in a small confocal volume. Therefore, FCS requires a fluorophore and FCCS, two non-overlapping fluorescent labels. Given these restrictions, FCS can almost be considered as a single solute sizing method.

Diffusion Ordered NMR SpectroscopY (DOSY) also yields the translational diffusion coefficient D but requires relatively long acquisition time. Moreover, the low sensitivity can become critical in the case of nanoparticles [18,19].

In this context, Taylor's dispersion analysis emerges as an interesting alternative, which allows for the calculation of hydrodynamic radii, based on measurements of diffusion coefficients. The capillary format of this method has really boosted its use in the pharmaceutical field, given the small sample volumes required. Hawe et al. compared the use of DLS and TDA in the field of therapeutic proteins [20]. They demonstrated that these two techniques have an area of overlap for several applications. Nevertheless, they are complementary by their very nature. This complementarity is particularly powerful in aggregation studies. TDA is a calibration-free sizing method, which is very appropriate in the low size range, and therefore complementary to DLS. Moreover, as the detection signal is mass-weighted, size distributions are not biased towards the largest sizes. On the contrary, DLS is very sensitive to these latter. By design, one of the key advantages of TDA—and by no means the least—is its inherent ability to be implemented with a wide range of detection methods, which allow for the adjustment of its sensitivity and selectivity.

## 5. Is There a "Universal" Detector?

### *5.1. UV, a Readily Useable Detection*

UV detection has been widely used in TDA for measuring hydrodynamic radii of a wide variety of molecules and nano-objects. One of the main reasons for this widespread use is undoubtedly the fact that commercially available capillary electrophoresis instruments equipped with UV detectors can be easily and directly used in TDA. Some authors also report the possibility of on-line measurement using recovery in a UV cuvette and cumulative detection [21].

The number of applications documented in the literature is substantial, and only a few representative examples are listed in Table 1. These papers demonstrate the value of this technique in many application fields where size matters: polymers [22], peptides/proteins/antigens [23-25], amyloid aggregates [26,27], DNA complexes [28], nanoparticles [29–31], nanocarriers [32], contrast agents [33], etc.

**Table 1.** Overview of the conditions used in TDA-UV for various applications.

| Sample | Concentration | Hydrodynamic Diameter | Mobilization Medium | $\lambda_{UV}$ | Injection Mode | Reference |
|---|---|---|---|---|---|---|
| poly(acrylic acid-co-acrylamide) standards | 0.22–2 g/L in 20 mM Tris/11 mM HCl with 35.5 mM LiCl | 20 to 380 nm | Sample buffer | 200 nm | Plug injection | [22] |
| Peptides | >0.5 mg/mL in 10 mM ammonium acetate buffer, pH 4.5. or 10 mM Tris-Gly buffer, pH 8.3 | | Sample buffer | 214 nm (for concentrations below 5 mg/mL) or 280 nm | Plug injection | [23] |
| Collagen | 3.65 g/L in 10 mM Tris, pH 7.4 | 2.3 nm and 7.4 nm | | 200 nm | Frontal mode | [24] |
| Vaccine antigens | 0.4 g/L in different analytical buffers | 4 nm, 5.5 nm, 13 nm, 34 nm | Sample buffer | 215 nm | Plug injection | [25] |
| β-amyloid peptides aggregates | 100 μM in 20 mM phosphate buffer pH 7.4 | 1–300 nm | Sample buffer | 191 nm | Plug injection | [26] |
| Aβ1-42 peptide | 100 μM in 10 mM Tris, 15 mMNaCl, pH 7.4 | 1–10 nm | Sample buffer | 200 nm | Plug injection | [27] |
| Polyplexes | 0.1 g/L for DNA and 0.6 g/L (0.8 g/L) for the PLKC (DGL) polycation in 10 mM Tris, HCl at pH 7.4 | 50–80 nm | | 200 and 260 nm for poly-L-lysine-base complexes 214 nm for dendrigraft-based complexes | Frontal mode | [28] |
| Nanolatexes | ca. 2.5 g/L in 12.7 mM sodium borate, pH 9.2 | 100–140 nm | Sample buffer | 214 nm | Plug injection | [29] |
| SPION nanoparticles (stabilized by citrate) | n.s. in MilliQ-water | 10–25 nm | Water | n.s. * | Plug injection | [30] |
| Silica nanoparticles | 2 g/L in water | 25 nm, 70 nm, 250 nm | Water | 214 nm | Plug injection | [31] |
| Gold nanoparticles (stabilized by citrate) | ca. 100 mg/L in water | 3.5 nm, 25 nm, 70 nm | 0.001% (*w*/*v*) TWEEN®20 | 520 nm | Plug injection | [31] |
| Lipid nanoparticles encapsulating mRNA | n.s. in10 mM PBS (phosphate-buffered saline), pH 7.4 | 86 nm | Sample buffer | 200 nm | Plug injection | [32] |
| Gd based contrast agents | 5 mM in10 mM Tris, pH 7.4 or pure water | 1–2.2 nm | | 200 nm | Frontal mode | [33] |

* n.s.: not specified.

The small class of molecules exhibiting a significant UV absorbance is one of the main limitations plaguing the use of TDA with UV. As a result, applications are often dedicated to measuring the hydrodynamic radii of standard compounds in solution in well-characterized buffers to avoid buffer mismatches [34]. The concentrations used are also relatively high. Several options have been considered to improve detection limits. One of the main approaches is to conduct TDA experiments in frontal mode (instead of plug injection) to improve detection sensitivity by avoiding the effect of dilution [24,28,33,35,36]. Malburet et al. [25] also showed that the UV detection limit is a linear function of $1/\sqrt{t_0}$. Consequently, the lower the flowrate, the more sensitive the method. However, this solution results in only a slight reduction in detection limits at the expense of analysis time.

A third solution involves on-line derivatization of the compounds to impart a specific absorbance to them. Thus, UV detection of carbohydrates was performed using TDA-UV. [37]. High pH (~12–13) of background electrolyte was, however, required to produce UV-absorbing intermediate species by photo-oxidation under UV radiations at the detection point.

### *5.2. Solutions for Non-UV Absorbing Solutes*

#### 5.2.1. Backscattering Interferometry (BSI)

One of the easiest ways to overcome this problem is to perform refractive index (RI) measurements. RI detectors are reasonably sensitive, providing a signal for essentially all analytes.

RI detection coupled with Taylor dispersion has long been reported for the measurement of diffusion coefficients of salts and mutual diffusion coefficients of solvent mixtures [38–40]. In this case, the experimental setups differ greatly from those currently used in analysis. They involve much longer tubes, resulting in very long analysis times and diameters approximately 10 to 100 times larger. However, since the refractive index is a property of the bulk solution, and not a property specific to a solute, measuring particle size in mixtures presents a number of problems. This type of detection does not appear suitable for TDA in capillary tubes, and backscattering interferometry (BSI) is preferred instead, as it provides access to variations in refractive indices via an optical measurement. As with UV detection or fluorescence, that will be discussed later, this detection method can therefore be implemented directly "on-capillary."

In the case of BSI detection, the RI increment with the concentration is the main parameter that controls the sensitivity of detection. Polysaccharides, which are typically non UV-absorbing compounds, were successfully analyzed and detected by TDA-BSI. However, sensitivity of detection is relatively low. No limits of detection (LOD) were reported using classical plug injections, but the authors worked with 20 g/L solutions [37]. The frontal mode used to enhance the sensitivity resulted in LOD ≈ 50–80 mg/L [41].

#### 5.2.2. Surface Plasmon Resonance (SPR)

Recently, D-SPR has emerged as a new method, where D stands for Diffusion [42–47]. Its case is somewhat different because the measurements are performed from commercially available SPR instruments [42], by only increasing the introduction capillary, in which Taylor diffusion will then occur. The SPR-based detector coupled with the Taylor dispersion injection was reported to determine diffusion coefficients of biomolecules: insulin [43], albumin [44], but also graphene dots, with higher hydrodynamic radii (ca. 12 nm) [45].

Since, SPR is capable of detecting changes in RI occurring near the surface of a sensor chip, the method is considered as label-free. However, according to Saetear et al. [41], the SPR signal is interrogating the liquid at about 100–200 nm from the surface, which cannot be considered as a true bulk measurement.

#### 5.2.3. Conductimetric Detection (C4D)

To the best of our knowledge, only a few publications have examined the coupling between TDA and C4D [48,49]. C4D is particularly well-suited for ionizable species that do not absorb UV light. Unlike with optical detections, these detection limits do not appear to be affected by a reduction in capillary diameter.

Somnin et al. [48] demonstrated that the sensitivity of this detector increases with charge density but reaches a plateau due to the condensation of counter-ions. Thus, a polymer chain will have an effective charge lower than the nominal charge due to the ionized monomers. They thereby stated that, in practice, the applicability of TDA-C4D is limited to polyelectrolytes with a molar mass below $10^5$ g/mol.

These combinations make it possible to analyze a large number of compounds, provided that the concentrations used are generally high (or in even in frontal mode), and that the analysis is performed to neat buffers.

## 6. Looking for Detection Selectivity

### *6.1. Fluorescence, an Optical Detector That Combines Sensitivity and Specificity*

Fluorescence is undoubtedly one of the most sensitive optical detection methods. Commercial devices designed for TDA are now equipped with this type of detection. However, it requires either fluorescent samples or fluorescent labeling. As a result, extending the application range implies that size measurements are often constrained measurements of labeled compounds with fluorescent tags, which may modify the genuine size.

The primary application of fluorescence-based TDA is not the measurement of hydrodynamic radii per se. Nevertheless, there are a few examples that make use of either the intrinsic fluorescence of proteins [50], or the use of fluorescent probes [51]. In the latter case, the compounds or objects being measured are much larger than the probe. However, the use of a probe is not always harmless. For example, Deleanu et al. demonstrated that the presence of a FITC tag on amyloid peptides alters their aggregation process [52]. Obeid et al., meanwhile, showed through analyses of extracellular vesicles (EVs) that the labeling procedure itself can lead to changes in size [53].

Unlike the detection methods described earlier, the high selectivity of fluorescence allows for the use of mobilization media that are more complex than simple buffers. As a result, this coupling is well-suited for the study of rapid biomolecular interactions, and is gradually becoming a pharmaceutical method in its own right, used routinely to explore biomolecular interactions.

It has numerous applications in the quantification of biological markers—such as serum proteins (e.g., human serum albumin) and antibodies [54–56]—as well as in the determination of affinity constants [51,57,58], and the formation of protein corona around nanoparticles [59]. Due to its robustness, measurements can be performed in media as complex as plasma, diluted serum [51], and clarified fermentation broths [54]. One study even shows that the use of 85% serum has little or no effect on the accuracy of results [56].

All these studies demonstrate the remarkable potential of this coupling, provided a suitable fluorescent probe is available.

*6.2. ICP-MS, an Attractive Detector, Both Selective and Multi-Elemental*

The TDA-ICP-MS coupling was considered a label-free method, but is of course limited to metal-containing compounds or nanoparticles. Although many potential applications have been anticipated, this hyphenation has so far been used only for the analysis of nanoparticles in a few papers [60–63], and this coupling remains to date highly confidential. One possible explanation is the lack of commercially available couplings, with the interfaces described in the literature being home-made sheath-flow interfaces. A very good correlation has been demonstrated between measurements obtained using UV-TDA and ICP-MS-TDA [60], indicating that these interfaces do not introduce any significant additional sources of dispersion.

The robustness of ICP-MS coupled to the use of a sheath liquid helps to overcome potential matrix effects of the various biological media. Thus, gadolinium-based nanoparticle size measurements in various media (serum, urine, cerebrospinal fluid) were performed without dilution [60]. In the same way, analyses of gold nanoparticles were performed in buffered solutions containing proteins up to 50 mg/mL to assess protein corona formation [61].

*6.3. Has Mass Spectrometry (MS) Underused Assets?*

Several papers dealing with TDA-MS can be found in the literature. This coupling seems very attractive, offering both selectivity and sensitivity without labeling, but its greatest advantage is undoubtedly the ability to perform multiplexed measurements. Two main interfaces were considered: (i) the direct coupling of the capillary to a mass spectrometer [21,64–69], and (ii) the use of a sheath-flow interface to improve sensitivity and stability of the electrospray–mass spectrometry (ESI-MS) signal [70–74].

Very few studies have reported size measurements of proteins using this coupling [67–69]. This is likely attributed to the ion suppression effect occurring in ESI-MS, which causes signal suppression at the center of the peak. Eisert et al. [68] proposed the introduction of a masking step in the fitting procedure and showed a very good correlation (0.99) between the hydrodynamic radii measured by such a coupling and those measured using TDA-fluorescence, across a range of proteins with radii ranging from 0.5 to 5 nm.

The benefits of using TDA-ESI-MS can be illustrated by the study of the oligomerization states of concanavalin A [69]. As measurements based on fluorescence detection provide only an averaged radius, hydrodynamic radii of both dimer and tetramer were independently measured by TDA-ESI-MS.

TDA-ESI-MS was also used to measure interactions in solutions. In this case, the stability of the gas-phase complexes is no longer a constraint, compared to ESI-MS. Clark et al. [72] reported the use of a make-up solvent to provide highly denaturating conditions just before the detection, thus enabling independent detection of the two entities.

Thanks to the multiplexing ability of this coupling, binding experiments can be performed using mixtures of ligands, as illustrated by the screening of different saccharides for non-covalent binding to lysozyme [74].

In those cases, buffers were always classical MS-compatible solutions such as ammonium acetate and water/acetonitrile mixtures. For this reason, some papers published using this coupling diverted TDA from its original purpose to develop a method for the online desalting of proteins and oligonucleotides [64,65,70,75]. By taking the advantage of the difference in diffusion coefficients between salts/small molecules and proteins, they showed the possibility of obtaining clear MS spectra directly from solutions of proteins in PBS [64] or antibodies in their formulation medium [70]. TDA is, however, relegated to the role of a separative method.

Interestingly, this property was exploited for the screening of libraries of 26 ligands of proteins [64], with computer--assisted targeted/non-targeted approaches [75], thus paving the way for high-throughput screening. This methodology is of course still restricted to ligands with moderate-to-high affinity constants.

## 7. What About Multi-Detection?

Several studies have examined the combination of multiple detectors, either at different points along the capillary [76–79] or at a single detection point [49].

If by "multi-detection", we mean "multi-point detection", its contribution is undeniable for improving the accuracy of hydrodynamic radius measurements. Cottet et al. showed that this approach eliminates errors arising from the injection volume, the pressure ramp, and discontinuities in the flow velocity [76]. It becomes essential if large sample volumes are injected. Technically, such a device has already been developed for UV applications. Recently, advances—particularly in 3D printing—have made it possible to implement dual fluorescence detection that is fully integrated into commercial capillary electrophoresis systems with detection limits of 600 pM for fluorescein [77–79]. This makes it a very accurate tool. However, Casto et al. reported a dramatic reduction in sensitivity due to the use of LED instead of laser to induce fluorescence [77].

If "multi-detection" means "detections using several detectors", it becomes a real asset for improving information. In this context, the multi-detection associated with TDA becomes a real strategy for the acquisition of simultaneous orthogonal information.

In practice, it is possible to couple many detectors with TDA simultaneously. Zeng et al. demonstrated the feasibility of triple detection (UV, fluorescence, and C4D) using a single detection window. They demonstrated the value of this system for analyzing a mixture containing fluorescent compounds, as well as both UV- and non-UV-absorbing compounds [49]. The combination of in-line optical detection and on-line mass spectrometry has also demonstrated its feasibility but has never been used to take advantage of the complementary nature of the information that the different Taylorgrams could provide [66].

Labied et al. showed that using when TDA-ICP-MS, the erosion of silica nanoparticles bearing Gd-chelates at the surface no longer allows us to track the nanoparticle [62]. The same problem was mentioned by Pedersen et al. in the study of HSA unfolding. The release of fluorescein from the HSA precludes any monitoring by fluorescence. They combined UV and fluorescence modes in TDA and showed, thanks to UV, that unfolding begins at 4 M urea. On the other side, the release of fluorescein from the HSA complex indicates that the binding site in HSA is undergoing local changes [80]. Malburet et al. proposed the on-line coupling of TDA with both UV and fluorescence detectors to provide complementary information on the encapsulation of mRNA in lipid nanoparticles [81]. By carefully selecting a fluorescent probe that does not penetrate the nanoparticles, they were able to determine the percentage of both encapsulated mRNA and degraded mRNA by ribonucleases.

## 8. Analytical Purposes Related to the Detection Mode Used in the TDA

Although many detection methods have been implemented with TDA, UV and fluorescence detection appear to account for the lion's share in this field (Table 2). Due to their respective features, TDA-UV studies generally aim to determine hydrodynamic radii in neat buffers, without labeling. Therefore, it demonstrates its value in the quality control of batches of therapeutics during storage. However, the lack of selectivity in UV detection prevents analysis under the conditions in which these therapeutics are used (e.g., blood, when injected intravenously). Analyses in such complex biological media require more selective detection modes. The use of a fluorescence detector often involves a labeling step for the compound being tested. This derivatization is not feasible for actual batches. In addition, the addition of a fluorescent tag may lead to changes in size, conformation, or properties (e.g., aggregation). Nevertheless, TDA coupled with fluorescence proves particularly valuable in interaction studies.

Mass spectrometry-based hyphenations with TDA offer interesting possibilities due to their multiplexing potential. However, the constraints imposed by ESI-MS still limit applications to fundamental studies in MS-compatible solvents. ICP-MS coupling, on the other hand, due to its robustness, allows for analyses in biological matrices without any sample preparation. Unfortunately, it is limited to metal-containing compounds.

**Table 2.** Analytical purposes associated with the detection mode used in TDA.

| Detection Used with TDA | Solutes | Analytical Purpose | | Buffers | References |
|---|---|---|---|---|---|
| UV | Nearly all | Quality control: Sizing under storage conditions | Characterization | Common buffers: PBS, Tris, Bistris, Phosphate buffer | [28,32,33,81–83] |
| | | | Aggregation studies | Common buffers: PBS, physiological serum | [20,84–86] |
| | | | Drug release | PBS | [87] |
| | | Interaction studies | Affinity constant determination | Tris | [88] |
| | | Research in neurodegenerative diseases | Aggregation studies | Phosphate buffer | [26] |
| SPR | Nearly all | Quality control: Sizing under storage conditions | Oligomeric forms | Common buffers: PBS, Tris | [43] |
| | | | Folding | PBS | [44] |
| Fluorescence | Fluorescent compounds or tagged compounds | Interactions studies | Affinity constant determination | Common buffers: PBS, HEPES or physiological serum)—0.1% bovine serum albumin may be added | [51,57,58,89–92] |
| | | | | Complex biological media: serum or diluted serum, fermentation broth | [51,54,57,93] |
| | | Quantification | | Plasma, serum | [53,92] |
| | | Quality control: Sizing under storage conditions | Characterization | | [94] |
| | | | Drug loading | PBS | [86] |
| | | | Stability | Simple buffer with guanidine hydrochloride | [50] |
| | | Quality control: Sizing under use conditions | Protein corona formation | PBS with unlabeled nanoparticles | [59] |
| | | | Binding heterogeneity | 85% plasma | [56] |
| | | Research in neurodegenerative diseases | Aggregation studies | Plasma | [95] |
| ESI-MS | Nearly all | Interactions studies | Ligand Screening | MS compatible buffers (mainly ammonium acetate) | [65,72,74,96] |
| ICP-MS | Metal-containing compounds | Quality control: Sizing under use conditions | Degradation in biological media | Serum, urine cerebrospinal fluid | [62] |
| | | | Protein corona formation | Buffers with high protein contents | [61] |

## 9. Conclusions and Future Directions

The biopharmaceutical industry still lacks effective methods for analyzing products under actual use conditions. Although solutions have been considered in TDA for specific classes of pharmaceutical compounds, such as highly fluorescent or metal-containing compounds, research in this area remains limited. Considerable further effort is needed to enable the analysis in more complex matrices. Combining TDA with other detectors could pave the way for new, robust and selective methods, and fill the gaps left by other techniques.

This powerful tool could benefit many other fields of application. Multi-detection could provide complementary information to improve our understanding of biological phenomena. For instance, the abnormal distribution of metals is of great interest in the study of degenerative diseases such as Parkinson's disease, Alzheimer's disease and amyotrophic lateral sclerosis. TDA coupled to a 'universal' detector could provide a macroscopic view of the different oligomerization states, which could be complemented by analyzing the distribution of metals between these forms using TDA-ICP-MS. However, the advantages of certain specific detection methods, such as analyzing low concentrations and complex mixtures, are negated if these detectors are used simultaneously.

**Supplementary**

**Author Contributions:** Conceptualization, A.H.; methodology, A.H.; writing—original draft preparation, R.L. and A.H.; writing—review and editing, R.L. and A.H.; visualization, R.L. and A.H.; supervision, A.H.; project administration, A.H.; funding acquisition, A.H. All authors have read and agreed to the published version of the manuscript.

**Funding:** This research was funded by a French Government Grant managed by the French National Research Agency (Grant ANR-22-CE44-0027-03).

**Institutional Review Board Statement:** Not applicable.

**Informed Consent Statement:** Not applicable.

**Data Availability Statement:** No new data were created or analyzed in this study. Data sharing is not applicable to this article.

**Acknowledgments:** During the preparation of this manuscript, the authors used the free version of DeepL for the purposes of translation. The graphical abstract was modified from a first version elaborated by GPT-5.6 Luna. The authors have reviewed and edited the output and take full responsibility for the content of this publication.

**Conflicts of Interest:** The authors declare no conflicts of interest.

## Abbreviations

The following abbreviations are used in this manuscript:

| | |
|---|---|
| AF4 | Asymmetric Flow field Flow Fractionation |
| BSI | Backs~~S~~cattering Interferometry |
| C4D | Capacitively ~~-~~Coupled Contactless Conductivity |
| DLS | Dynamic Light scattering |
| DOSY | Diffusion Ordered NMR SpectroscopY |
| ESI-MS | Electrospray Ionization–~~—~~Mass Spectrometry |
| EV | Extracellular Vesicle |
| FCS | Fluorescence Correlation Spectroscopy |
| FCCS | Fluorescence Cross-Correlation Spectroscopy |
| FIDA | Flow Induced Dispersion Analysis |
| ICP-MS | Inductively Coupled Plasma-—Mass Spectrometry |
| LOD | Limit Of Detection |
| MS | Mass Spectrometry |
| NMR | Nuclear Magnetic Resonance |
| NTA | Nanoparticle Tracking Analysis |
| PBS | Phosphate--Buffered Saline |
| SEC | Size Exclusion Chromatography |
| SPR | Ssurface Plasmon Rresonance |
| TDA | Taylor Dispersion Analysis |
| TEM | Transmission Electron Microscopy |
| UV | Ultra-Violet |